\documentclass[preprint,%twocolumn,%
prb,aps,floats,amssymb,showpacs,floatfix,superscriptaddress]{revtex4}
\usepackage{graphicx,bm}

\newcommand\Tr{\mathop{\rm Tr}\nolimits}

\makeatletter
\def\graphicscale{\twocolumn@sw{0.33}{0.5}}
\makeatother

\begin{document}

\date{July 30, 2007}

\title{
Evidence for a floating phase of the \\
transverse ANNNI model at high frustration 
}

\author{Matteo Beccaria}
    \email{matteo.beccaria@le.infn.it}
    \affiliation{Dipartimento di Fisica dell'Universit\`a di Lecce
        and I.N.F.N., Sezione di Lecce, 
        Via Arnesano, 73100 Lecce, Italy}
\author{Massimo Campostrini}
    \email{massimo.campostrini@df.unipi.it}
    \affiliation{Dipartimento di Fisica dell'Universit\`a di Pisa
        and I.N.F.N., Sezione di Pisa,
        Largo Bruno Pontecorvo 3, 56127 Pisa, Italy}
\author{Alessandra Feo}
    \email{feo@fis.unipr.it}
    \affiliation{Dipartimento di Fisica dell'Universit\`a di Parma
        and I.N.F.N., Gruppo coll.\ di Parma,
        Viale G. P. Usberti 7/A, 43100 Parma, Italy}

\begin{abstract}
We study the transverse quantum ANNNI model in the region of high
frustration ($\kappa>0.5$) using the DMRG algorithm.  We obtain a
precise determination of the phase diagram, showing clear evidence for
the existence of a floating phase, separated from the paramagnetic
modulated phase by a high-order critical line ending at the
multicritical point.  We obtain simple and accurate formulae for the
two critical lines.
\end{abstract}


\maketitle

\section{The model}
\label{sec:model}

The ANNNI model is an axial Ising model with competing
next-nearest-neighbor antiferromagnetic coupling in one direction.  It
is a paradigm for the study of competition between magnetic ordering,
frustration and thermal disordering effects.

In the Hamiltonian limit, we consider a one-dimensional quantum spin
$S={\textstyle{1\over2}}$ chain interacting with an external magnetic
field, called the TAM model (transverse ANNNI).

The TAM Hamiltonian for $L$ spins with open boundary conditions
reads\cite{Selke,Beccaria:2005nz}
\begin{equation}
H = - J_1 \sum_{i=1}^{L-1} \sigma_i^z\sigma_{i+1}^z
    - J_2 \sum_{i=1}^{L-2} \sigma_i^z\sigma_{i+2}^z
    - B \sum_{i=1}^L \sigma_i^x . 
\label{eq:H-TAM}
\end{equation}
We use the ``traditional'' notation $\kappa = -J_2/J_1$.  The
notations $\lambda=J_1/B$ and $\Gamma=B$ are sometimes used in the
literature.

The sign of $J_1$ is immaterial, since the Hamiltonian is invariant
under the transformation
\begin{equation}
J_1 \to -J_1, \quad \sigma_i^y \to (-1)^i\sigma_i^y, 
              \quad \sigma_i^z \to (-1)^i\sigma_i^z.
\end{equation}
Likewise, the sign of $B$ is immaterial.  Without loss of generality,
we set $J_1=1$.  We restrict ourselves to positive $\kappa$ and even
$L$.

We also consider fixed boundary conditions, where we add to the
extremities of the chain two fixed spins $\sigma_0$ and
$\sigma_{L+1}$, with the possibilities of parallel
($\sigma_0,\sigma_{L+1}=\uparrow\uparrow$) or antiparallel
($\sigma_0,\sigma_{L+1}=\uparrow\downarrow$) boundary conditions.

In the region of high frustration ($\kappa>0.5$), despite extensive
studies\cite{Duxbury-Barber,Arizmendi-etal,Sen-etal,Guimaraes-etal,Chandra-Dasgupta},
the phase diagram of the transverse ANNNI model is not well known.
For low $B$, the model is known to be in the gapless ``antiphase''
$\uparrow\uparrow\downarrow\downarrow$.  It undergoes a second-order
phase transition at a magnetic field $B_1(\kappa)$.  The existence of
a ``floating'' phase, massless and with slowly decaying spin
correlation functions, up to a Kosterliz-Thouless phase transition at
a magnetic field $B_2(\kappa)$, is an open question.  For high $B$,
the TAM is known to be in a paramagnetic modulated phase.

\section{Observables}
\label{sec:observables}

We measure the two lowest energies $E_0$ and $E_1$, the mass gap
$\Delta=E_1-E_0$, the entanglement entropy $S_A$ (see below), and two
spin-spin correlation functions: the ``slow'' correlation function
\begin{equation}
c_s(d) = \langle \sigma^z_{L/2+1}\,\sigma^z_{L/2+1+d}\rangle,
    \quad 1\le d\le L/2
\label{eq:slow-correl}
\end{equation}
and the ``fast'' correlation function%
\footnote{The labels ``slow'' and ``fast'' only refer to the different
distances involved and have no physical meaning.}
\begin{equation}
c_f(d) = \langle \sigma^z_{L/2-d}\,\sigma^z_{L/2+1+d}\rangle,
    \quad 0\le d\le L/2.
\label{eq:fast-correl}
\end{equation}

Interesting quantities related to the correlation functions are:\\
the overlap $o$ of $c_s(d)$ with the antiphase correlation
function 
\begin{equation}
c_a(d) = (-1)^{\lfloor(d-L/2)/2\rfloor}, \qquad
o = {2\over L} \sum_{d=1}^{L/2} c_s(d)\,c_a(d);
\label{eq:antiphase}
\end{equation}
the average fast correlation function (times an oscillating sign)
\begin{equation}
\overline c_f = (-1)^{L/2} {2\over L+2} \sum_{d=0}^{L/2} c_f(d);
\label{eq:avg_fsc}
\end{equation}
the range of the fast correlation function
\begin{equation}
R = {\sum_{d=0}^{L/2} d\,c_f^2(d) \over \sum_{d=0}^{L/2} c_f^2(d)}.
\label{eq:range_fsc}
\end{equation}

\subsection{Entanglement entropy}
\label{sec:entanglement-entropy-def}

It is possible to study an order-disorder phase transition using the
entanglement entropy\cite{Calabrese:2004eu}.  We divide the system of
size $L$ into a left subsystem of size $\ell$ and a right subsystem of
size $L-\ell$, and define
\begin{equation}
S_A(\ell;L) = - \Tr(\rho_A \ln \rho_A),
\end{equation}
where $A$ denotes the degrees of freedom of the left subsystem, $B$
the degrees of freedom of the right subsystem, and 
$\rho_A = \Tr_B|\Psi_0\rangle\langle\Psi_0|$; note that
$S_A(\ell;L)=S_A(L{-}\ell;L)$.  For a critical system we expect
(neglecting lattice artifacts)
\begin{equation}
S_A(\ell;L)\sim (c/6) \log(L\sin(\pi\ell/L)),
\label{eq:SA-crit-fv}
\end{equation}
where $c$ is the conformal anomaly number (central charge) of the
corresponding conformal field theory, and $\sim$ means ``up to a
(non-universal) additive constant''; for the case of interest for the
infinite-volume DMRG, $\ell={\textstyle{1\over2}} L$, and
$\sin(\pi\ell/L)$ only shifts the additive constant.  For a
noncritical system, we expect
\begin{equation}
S_A({\textstyle{1\over2}} L;L) \sim (c/6) [\log L + s(L/\xi)],
\label{eq:SA-FSS}
\end{equation}
where $s(x)$ is a {\em universal\/} finite-size scaling function
satisfying the constraints $s(0)=0$ and $s(x)\sim-\log x$ for large
$x$.

\subsection{Domain-wall energy}
\label{sec:DW-energy-def}

So far, we only considered open boundary conditions.  Following
Ref.~\onlinecite{Derian-Gendiar-Nishino}, we define the domain-wall
energy (note that our definition of $L$ differs by 2 from the
definition of Ref.~\onlinecite{Derian-Gendiar-Nishino})
\begin{equation}
E_{\rm DW}(\kappa,B,L) = 
(-1)^{L/2+1} \Bigl[E_0^{\uparrow\uparrow}(\kappa,B,L) - 
                 E_0^{\uparrow\downarrow}(\kappa,B,L)\Bigr],
\end{equation}
where $E_0^{\uparrow\uparrow}$ and $E_0^{\uparrow\downarrow}$ are the
ground state energies with parallel and antiparallel boundary
conditions respectively.

\section{The algorithm}
\label{sec:algorithm}

We implement the density matrix renormalization group (DMRG) algorithm
described in Ref.~\onlinecite{Schollwoeck-review}.  We sample the
$n_s$ lowest energy levels with equal weights, i.e., we use the
reduced density matrix
\begin{equation}
\hat\rho_S = 
{1\over n_s} \Tr_E \sum_{i=0}^{n_s-1} |\psi_i\rangle\langle\psi_i|
\label{eq:rho-hat}
\end{equation}
(see Eq.\ (26) of Ref.~\onlinecite{Schollwoeck-review}).  Usually,
since we are interested in the mass gap $\Delta$, we set $n_s=2$.  We
identify system and environment (for antiparallel boundary conditions,
up to a spin flip $\sigma_i^y \to -\sigma_i^y$, $\sigma_i^z \to
-\sigma_i^z$).  The typical dimensions of the truncated system and
environment $M$ range from $80$ to $160$; in the following, $M=80$
will be understood, unless $M$ is explicitly quoted.

The crucial part of the numerical computation is finding the lowest
eigenvalues and eigenvectors of the superblock Hamiltonian; we employ
the Implicitly Restarted Arnoldi Algorithm implemented in Arpack
\cite{ARPACK}, in the routine {\tt dsaupd} used in mode 1.  We use
(typically) 100 Lanczos vectors and require convergence to machine
precision, obtaining residual norms 
$|H x - \lambda x| / |\lambda| \sim 10^{-14}$.

We observe a truncated weight (the sum of the eigenvalues of the
density matrix whose eigenvectors are dropped in the truncation)
$\varepsilon\sim10^{-8}$ for ``normal'' configurations, and
$\varepsilon\sim10^{-7}$ for peaks of $\Delta$ (see below).

We managed to diagonalize the system exactly up to $L=22$; both the
finite- and the infinite-volume DMRG algorithm reproduce the results
of exact diagonalization.  For moderate $L$, finite- and the
infinite-volume DMRG give consistent results.  For higher $L$,
discrepancies between finite- and infinite-volume DMRG and
$M$-dependence of $\Delta$ becomes noticeable; they are strongly
observable-dependent, and they will be discussed below, where results
on observables are presented.

During a run of the finite-volume DMRG algorithm on a system with
$L_n$ sites, information about the system/environment for all smaller
system is available.  It is therefore possible, with a moderate extra
numerical effort, to estimate the observables for all the systems with
$L<L_n$ sites.  These estimates almost coincide with the results
obtained running independently at each $L$.  The additional errors
introduced by this procedure will also be discussed below.

The finite-volume DMRG algorithm at large $L$ requires a very large
amount of memory; however, since the observables at each lattice size
are accessed only twice per cycle, they can be conveniently kept on
disk, requiring only a very large amount of disk space; for $L=600$
and $M=80$, e.g., ca.\ 6 Gbytes are required.

\section{Phases at $\bm{\kappa=0.75}$}
\label{sec:phases-k=0.75}

We will first focus our attention on the model at $\kappa=0.75$, and
later on extend the study to other values of $\kappa$.

\begin{figure}[tbp]
\begin{center}
\leavevmode
\includegraphics*[scale=\graphicscale]{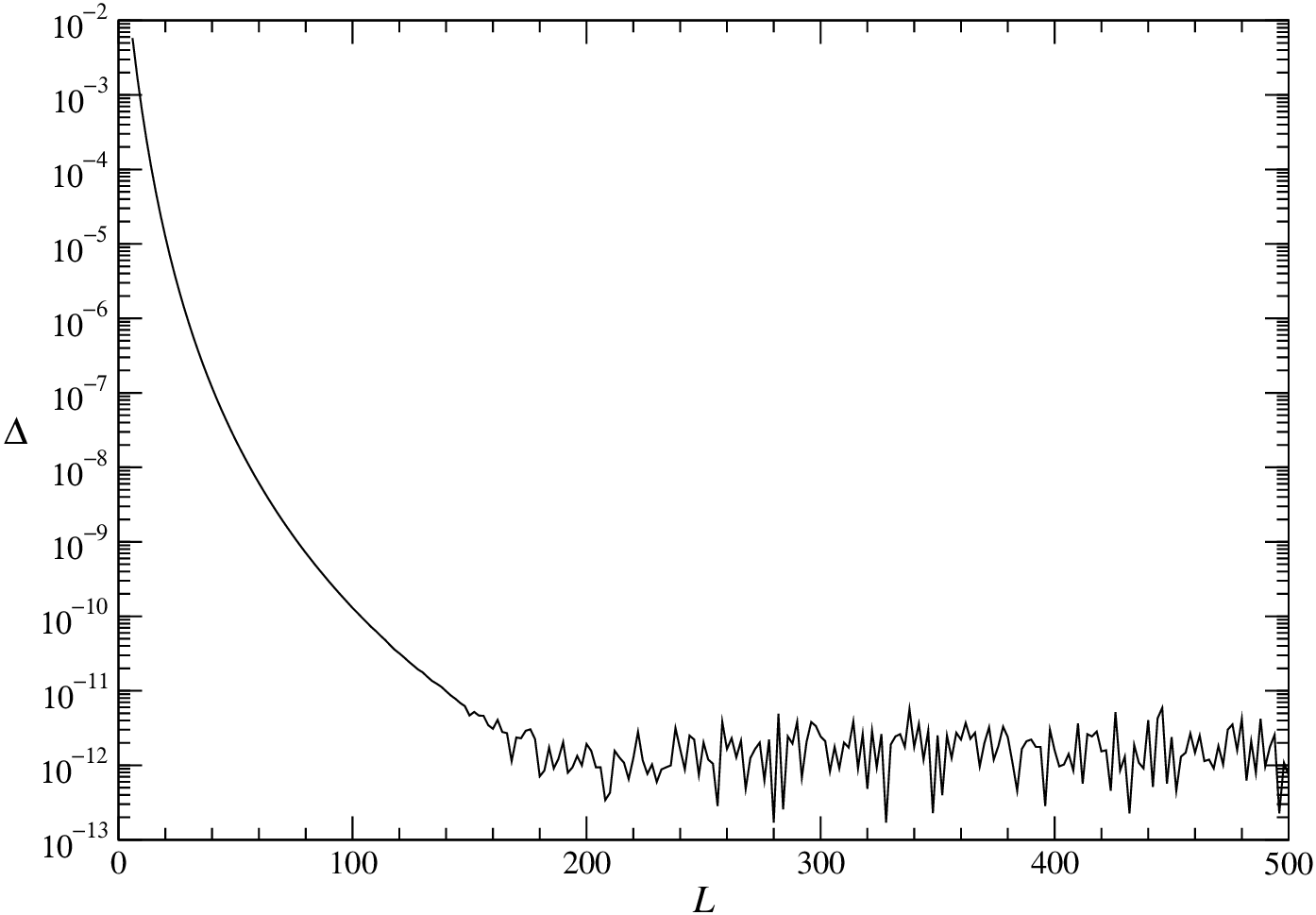}
\caption{Mass gap vs. $L$ for $\kappa=0.75$ and $B=0.257$.  To
appreciate the effect of numerical errors, note that
$E_0(B{=}0.257)\protect\cong-0.8L$.}
\label{fig:gap-B0.257}
\end{center}
\end{figure}

Running the infinite-volume DMRG algorithm at $\kappa=0.75$ and
$B\le0.257$, with open boundary conditions, we observe that
the mass gap $\Delta$ vanishes exponentially in $L$, apart
from numerical errors due to the fact that $\Delta$ is computed as
$E_1-E_0$: see, e.g.,\ Fig.~\ref{fig:gap-B0.257}.

The slow correlation function $c_s(d)$ almost coincides with
$c_a(d)$; the overlap $o$ approaches a value very close to 1 with
corrections proportional to $1/L$.
The fast correlation function $c_f(d)$ is constant and close to
$\pm1$, apart from $d=0$ and $d$s close to $L/2$;
the range $R$ is almost exactly $L/4$ (the value for a constant
$c_f(d)$) and the average $\overline c_f$ approaches a value very
close to $-1$ with corrections proportional to $1/L$.

There is a very sharp phase transition at $0.257<B_1<0.258$.  We will
postpone its detailed study, since it is best done using $E_{\rm DW}$.

\begin{figure}[tbp]
\begin{center}
\leavevmode
\includegraphics*[scale=\graphicscale]{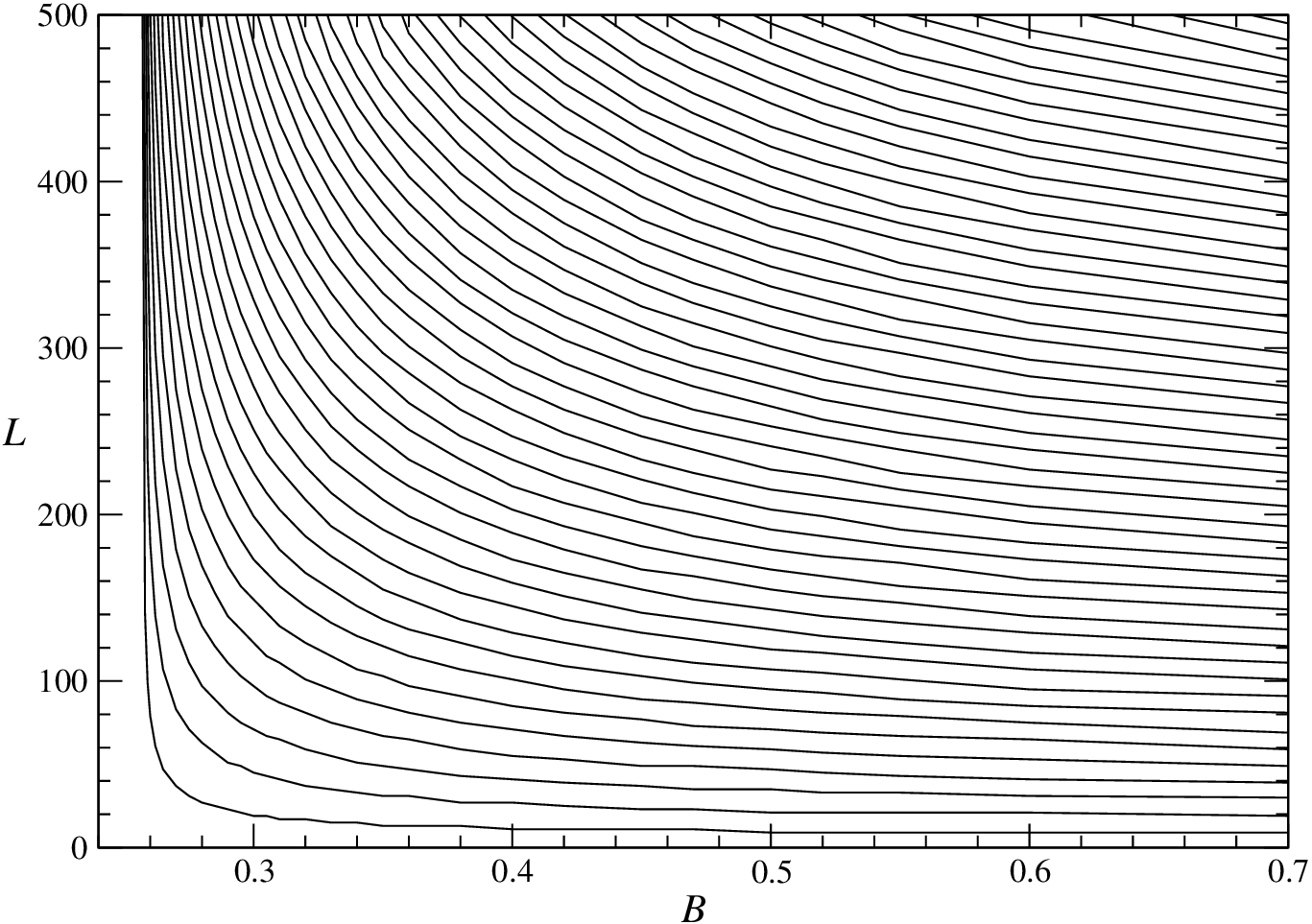}
\caption{Mass gap peaks in the $L$--$B$ plane for $\kappa=0.75$.}
\label{fig:L-B-plane}
\end{center}
\end{figure}

\begin{figure}[tbp]
\begin{center}
\leavevmode
\includegraphics*[scale=\graphicscale]{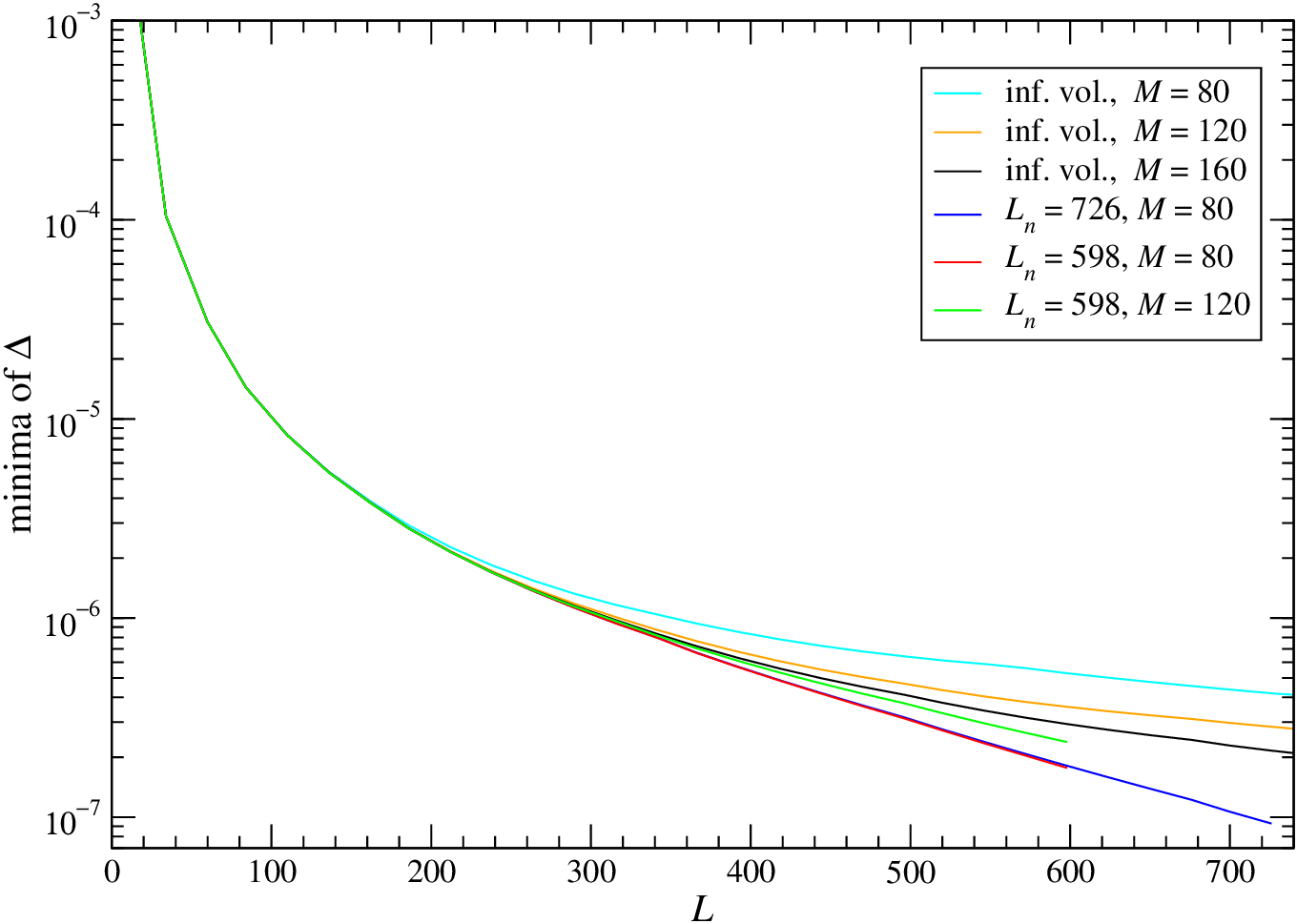}
\caption{Minima of mass gap $\Delta$ vs.\ $L$ for $\kappa=0.75$ and
$B=0.3$.}
\label{fig:gap_min-B0.3}
\end{center}
\end{figure}

Running at $\kappa=0.75$ and $B\ge0.258$, the mass gap $\Delta$ as a
function of $L$ at fixed $B$ shows sharp peaks with a frequency
increasing with $B$: see Fig.~\ref{fig:L-B-plane}.
Each peak match exactly a change of sign of $\overline c_f$.  At each
$B$, for $L$ smaller than the first peak we observe signals very
similar to the case $B\le0.257$; for higher $L$, we observe that the
minima of $\Delta$ seem to go to zero for $B\le0.4$ and to a nonzero
limit for $B\ge0.5$; however, the determination of $\Delta$ from the
infinite-volume DMRG is not accurate for $L\gtrsim200$; finite-volume
DMRG data with $M=80$ become unreliable for $L>300$; we
show in Fig.~\ref{fig:gap_min-B0.3} the case $B=0.3$.

We performed a finite-size analysis of $\Delta$, similar to the
analysis of Ref.~\onlinecite{Beccaria:2005nz}, but with a complication
arising from the peak structure.  We run the finite-volume DMRG
algorithm for $B=0.4$, 0.41, 0.42, 0.43, 0.44, 0.45, 0.46 and
$L_n\ge292$ corresponding to a minimum of $\Delta$.  For each $B$, we
select the minima of $\Delta$ and define $\Delta_L(B,\kappa)$ outside
the minima by interpolation in $L$.  We now take two values $L_1$ and
$L_2$ and look for the intersection $B_i(L_1,L_2)$ of the two curves
$L_1\Delta_{L_1}(B,\kappa)$ and $L_2\Delta_{L_2}(B,\kappa)$ vs.\ $B$
(interpolating in $B$ at fixed $L$ and $\kappa$ as needed).  The
results are shown in Fig.~\ref{fig:k=0.75,B_i}: we note that $M=120$
and $M=160$ data almost coincide, and even $M=80$ data are adequate in
the range of $L$s considered; we quote as a final result
$B_2=0.424(3)$.
\begin{figure}[tbp]
\begin{center}
\leavevmode
\includegraphics*[scale=\graphicscale]{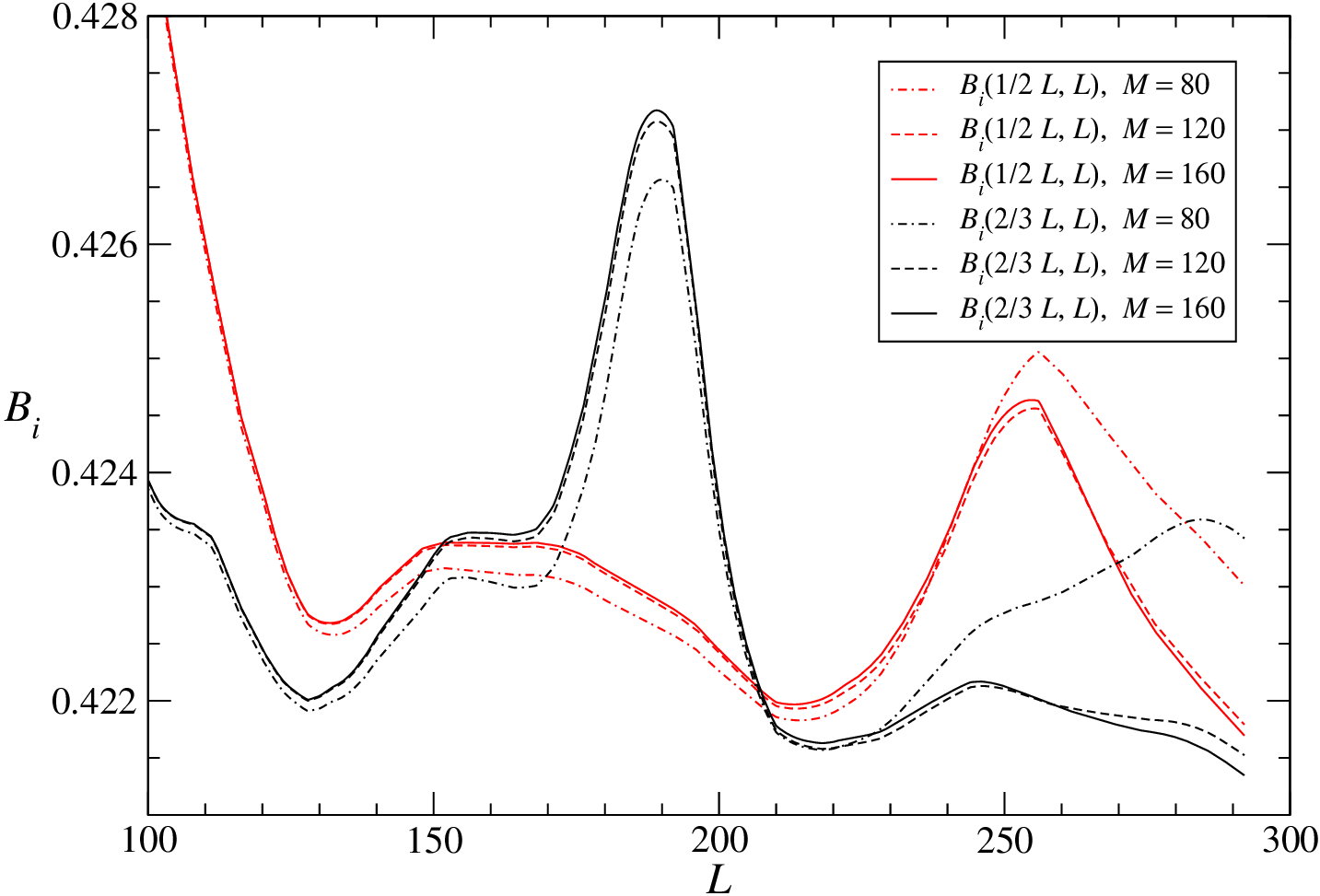}
\caption{The intersection $B_i$ 
vs.\ $L$, for $\kappa=0.75$.}
\label{fig:k=0.75,B_i}
\end{center}
\end{figure}
The data presented here were obtained from a run at a single $L_n$ for
each $B$ (see Sect.~\ref{sec:algorithm}); in order to check that the
error introduced is under control, we also performed separate runs for
all the values of $L$ required, for $N=80$ and for $N=120$ at
$B=0.42$, and repeated the analysis: the values of $B_i(L_1,L_2)$
never change by more than 0.0005.

\begin{figure}[tbp]
\begin{center}
\leavevmode
\includegraphics*[scale=\graphicscale]{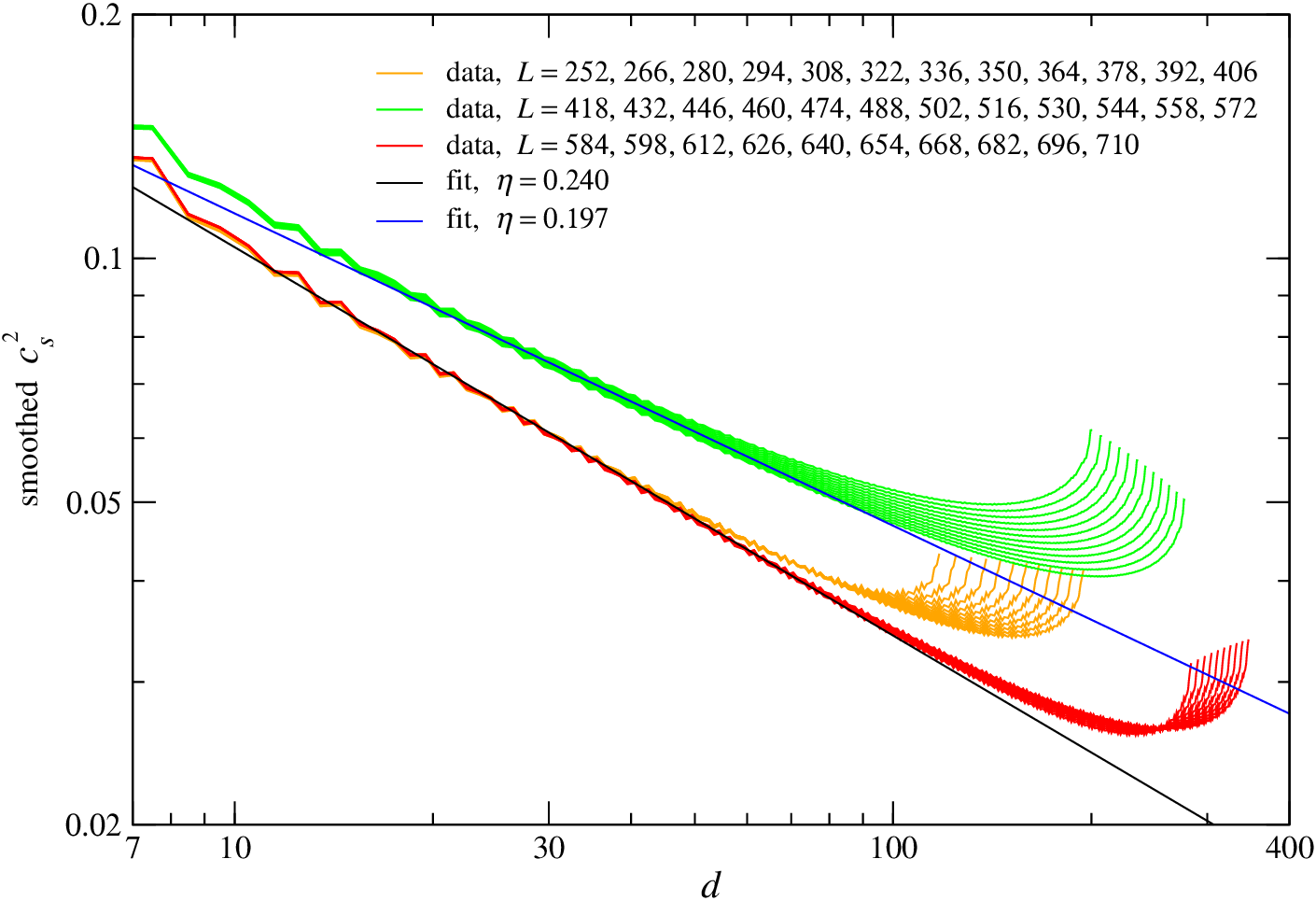}
\caption{The smoothed squared slow correlation function $c^2_s(d)$ for
$\kappa=0.75$ and $B=0.425$, from the finite-volume DMRG at $L_n=710$
and $M=120$, for values of $L$ corresponding to minima of $\Delta$.
(Data at $M=80$ give a very similar plot.)}
\label{fig:ssc2}
\end{center}
\end{figure}

For $0.258\le B\lesssim0.45$, the slow correlation function $c_s(d)$
at fixed $L$ shows oscillations with power-law damping, in rough
agreement with
\begin{equation}
c_s(d) \cong a d^{-\eta} \cos(qd + \phi),
\label{eq:cs-floating}
\end{equation}
with $\eta\sim0$ for $B=0.258$, increasing with $B$ but remaining
smaller than $\textstyle{1\over3}$.  For $B\gtrsim0.5$,
$c_s(d)$ at fixed $L$ shows oscillations with exponential damping.
We tried to extract $\eta$ by fitting $c_s^2(d)$, smoothed by taking a
running average over $\lfloor2\pi/q+{\textstyle{1\over2}}\rfloor$
points, to the form $a d^{-2\eta}$.  In Fig.~\ref{fig:ssc2} we show
the typical case $B=0.425$; $c_s^2(d)$ for different values of $L$
converge not to a single curve but to two separate curves, with
different values of $\eta$, preventing a precise determination of
$\eta$.

\begin{figure}[tbp]
\begin{center}
\leavevmode
\includegraphics*[scale=\graphicscale]{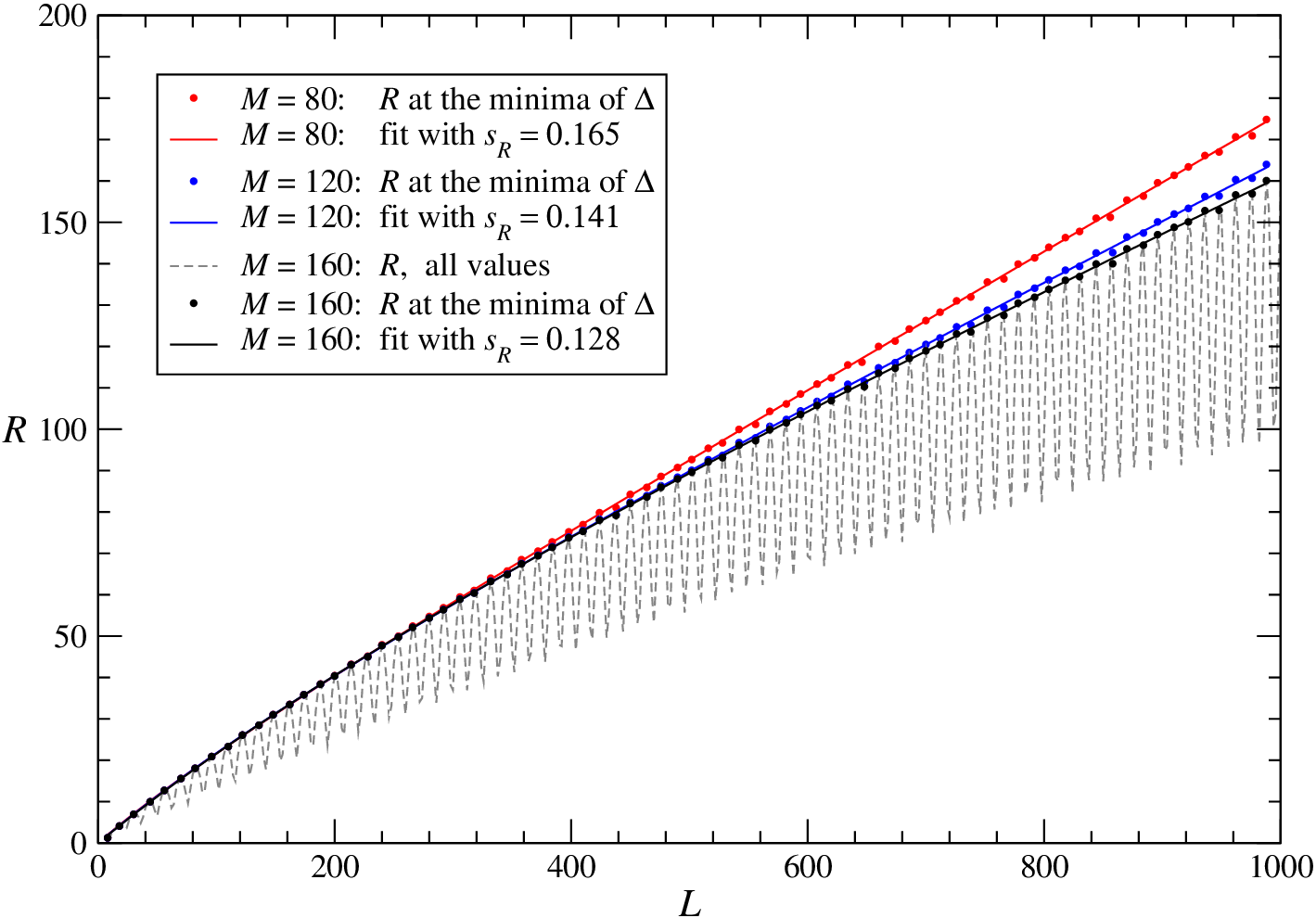}
\caption{$R$, at the $L$s corresponding to minima of $\Delta$, with
fits to Eq.\ (\ref{eq:s-R}), for $\kappa=0.75$, $B=0.45$ and different
$M$s.  For $M=160$, all the values of $R$ are also plotted.}
\label{fig:range-B0.45}
\end{center}
\end{figure}

The range of the fast correlation function $R$ should distinguish
clearly the floating phase, where $R\to\infty$ as $L\to\infty$ (since
$\eta<{\textstyle{1\over2}}$) , from the paramagnetic phase, where $R$
has a finite limit as $L\to\infty$.  A first problem is the presence
of oscillations, with dips corresponding to the peaks of $\Delta$ (see
Fig.~\ref{fig:range-B0.45}); it is solved by selecting the values of
$R$ at the $L$s corresponding to the peaks of $\Delta$.
After this operation, $R$ vs.\ $L$ at fixed $B$ and $M$ is well fitted
to the form
\begin{equation}
R(L) = {s_R L^2 + p L \over L+q} \; ;
\label{eq:s-R}
\end{equation}
if we plot the asymptotic slope $s_R$ vs.\ $B$ we should be able to
see a drop towards 0 in correspondence with the phase transition.
However, in the critical region, $R$ (and $s_R$) strongly depend on
$M$: see Fig.~\ref{fig:range-B0.45}; we can only conclude that
$B_2\simeq0.45$.

The analysis of the range of the slow correlation function gives
similar, but even less precise, results.  We conclude that the spin
correlation functions are unsuitable for the precise determination of
the floating/paramagnetic phase transition.

\subsection{Entanglement entropy}
\label{sec:entanglement-entropy-k=0.75}

We show in Fig.~\ref{fig:k=0.75,B=0.425,SA} the entanglement entropy
$S_A$ for the typical case $B=0.425$.  Finite-volume DMRG essentially
reproduces the results of infinite-volume DMRG at the same $M$.  We
can estimate that the DMRG determination of $S_A$ is reliable up to
$L=200$ for $M=80$, up to $L=300$ for $M=120$, and up to $L=400$ for
$M=160$.  For each value of $M$, within the given range of $L$, the
difference between finite- and infinite-volume DMRG results is less
than 0.001.  We can therefore compute $S_A$ using the faster
infinite-volume DMRG, and this allows us to work at larger values of
$M$.
\begin{figure}[tbp]
\begin{center}
\leavevmode
\includegraphics*[scale=\graphicscale]{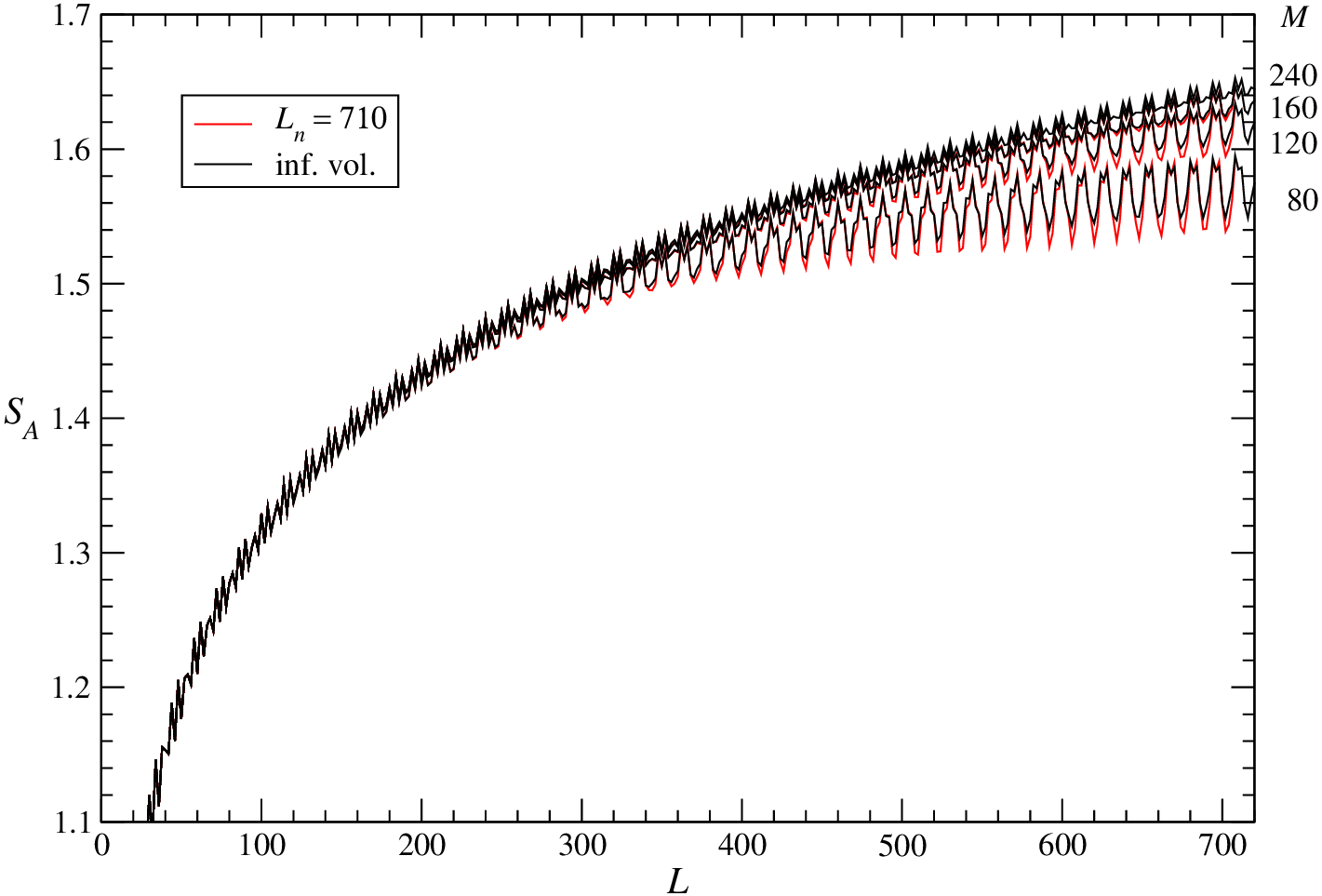}
\caption{The entanglement entropy
$S_A({\textstyle{1\over2}} L;L)$ vs.\ the size of the system $L$, for
$\kappa=0.75$, $B=0.425$, $L_n=710$, and infinite-volume DMRG.}
\label{fig:k=0.75,B=0.425,SA}
\end{center}
\end{figure}

In the antiphase, $S_A$ is essentially constant, indicating
a very small correlation length.  

The simple Ansatz for the finite-size scaling function entering Eq.\ 
(\ref{eq:SA-FSS}),
\begin{equation}
s(x) = -\ln(x + e^{-\alpha x})
\label{eq:s-FSS}
\end{equation}
with $\alpha\equiv1$, is found to fit the entanglement entropy data
very well (excluding just the very smallest lattices with
$L\lesssim10$) in all cases for the floating and paramagnetic phases;
$c$ is always compatible with 1.  The best determination of $\xi$,
obtained by fitting $S_A$ with $c\equiv1$ fixed, is shown in
Fig.~\ref{fig:k=0.75,xi-SA}; our final estimate is $B_2=0.44(1)$.
\begin{figure}[tbp]
\begin{center}
\leavevmode
\includegraphics*[scale=\graphicscale]{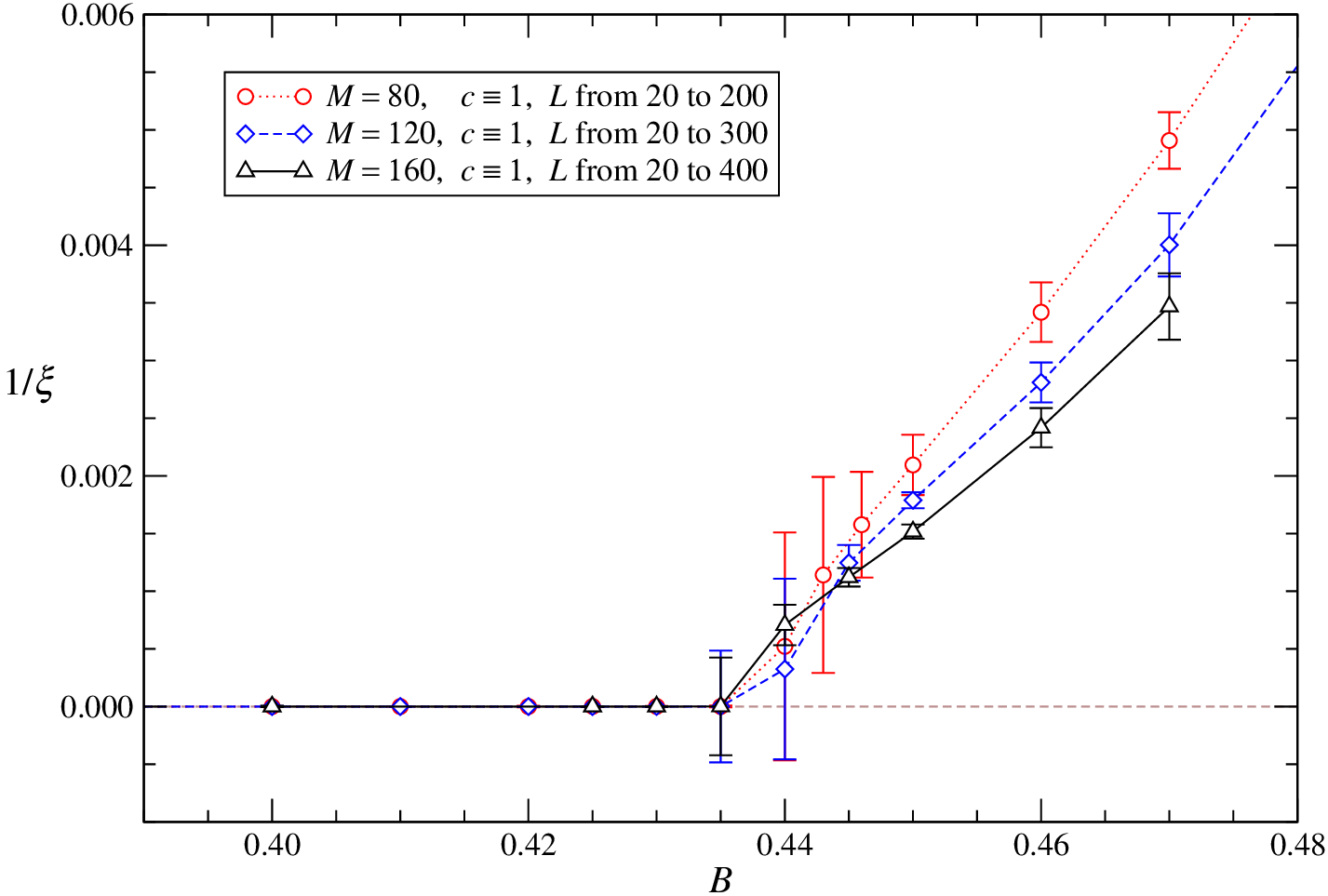}
\caption{The reciprocal correlation length $1/\xi$
(determined from $S_A$) vs.\ $B$, for $\kappa=0.75$.  Fitting the
$M=120$ and $M=160$ data for $20\le L\le200$ we obtain results almost
identical to the $M=80$ results plotted; likewise, fitting the $M=160$
data for $20\le L\le300$ we obtain results almost identical to the
$M=120$ results plotted; we omit these results from the plot for
readability.  All data for $B\le0.43$ are consistent with zero, with a
very small error ($\sim 10^{-6}$) which is not visible at the scale of
the plot.}
\label{fig:k=0.75,xi-SA}
\end{center}
\end{figure}

\subsection{Domain-wall energy}
\label{sec:DW-energy-k=0.75}

So far, we only considered open boundary conditions.  We now switch to
fixed boundary conditions, in order to compute the domain-wall energy
$E_{\rm DW}$; with fixed boundary conditions, there are no problems
with quasi-degenerate energy levels (typically, $\Delta>0.01$) or
peaks in $\Delta$ associated with level crossings, and truncated
weights are $\varepsilon\sim10^{-9}$ or smaller for $M=80$.  Even if
we are not interested in the mass gap, we run with $n_s=2$, which
gives results more stable than $n_s=1$.

We may fit $E_{\rm DW}$ to the form
\begin{equation}
E_{\rm DW} = a \exp(-dL) L^{-\nu} + E_\infty
\label{eq:DW-anti}
\end{equation}
in the antiphase 
and
\begin{equation}
E_{\rm DW} = {a \exp(-dL) \over L} 
\Bigl[|\cos(kL + \phi)| - |\sin(kL + \phi)|\Bigr]
\label{eq:DW-para}
\end{equation}
in the floating phase (with $d=0$) and in the paramagnetic
phase\cite{Derian-Gendiar-Nishino}.  The fits, excluding (typically)
lattices with $L<16$, are of very good quality and stable.

Eq.\ (\ref{eq:DW-anti}), fits perfectly $E_{\rm DW}$ for $B\le0.257$,
giving $E_\infty\to0$, constant $\nu\simeq1.6$ and $d\simeq0.008$ for
$B\nearrow B_1$.  Eq.\ (\ref{eq:DW-para}), with $d=0$, fits perfectly
$E_{\rm DW}$ for $B>0.257$ and gives $k\to0$ for $B\searrow B_1$.  The
best estimators of $B_1$ are $E_\infty$ in the antiphase and $k^2$ in
the floating phase, both vanishing linearly at $B_1$, see
Figs.~\ref{fig:k=0.75,Einf-EDW} and \ref{fig:k=0.75,k2-EDW}; the final
estimate of the critical field is $B_1=0.2574(2)$.
\begin{figure}[tbp]
\begin{center}
\leavevmode
\includegraphics*[scale=\graphicscale]{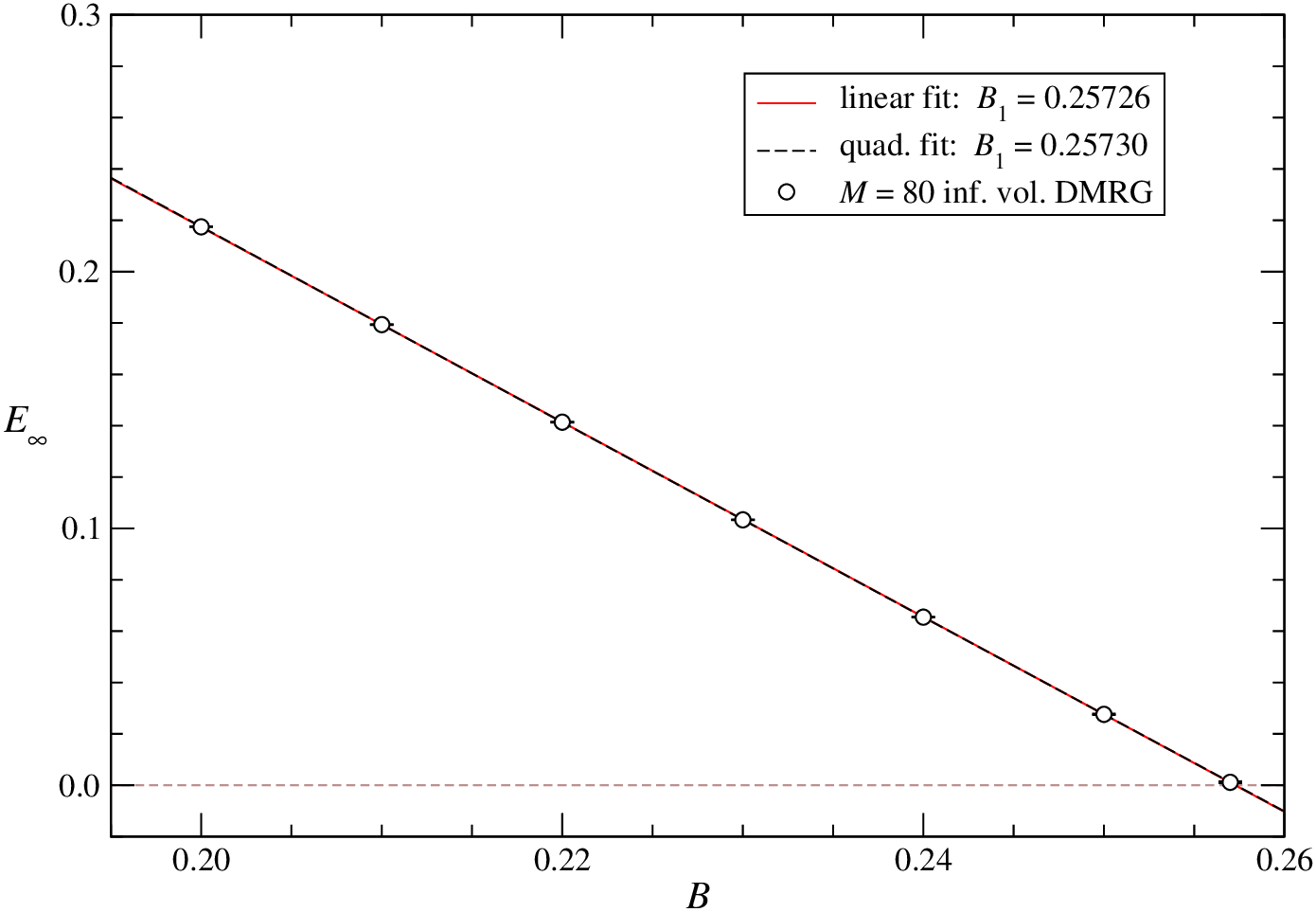}
\caption{The infinite-volume domain-wall energy 
$E_\infty$ vs.\ $B$, for $\kappa=0.75$.}
\label{fig:k=0.75,Einf-EDW}
\end{center}
\end{figure}
\begin{figure}[tbp]
\begin{center}
\leavevmode
\includegraphics*[scale=\graphicscale]{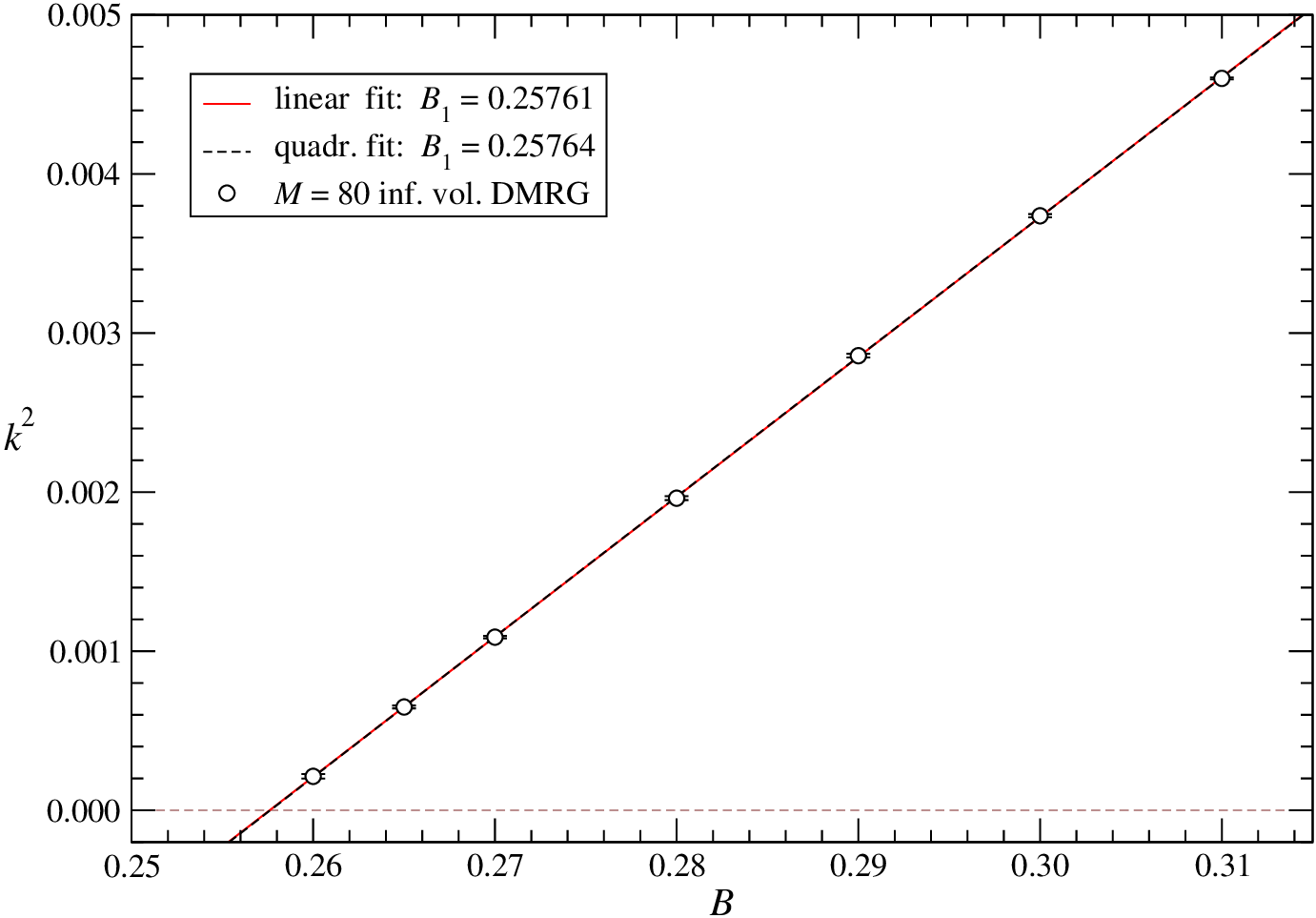}
\caption{The squared modulation parameter 
$k^2$ vs.\ $B$, for $\kappa=0.75$.}
\label{fig:k=0.75,k2-EDW}
\end{center}
\end{figure}

So far, we obtained results very similar to those of
Ref.~\onlinecite{Derian-Gendiar-Nishino}.  We turn now to the problem
of identifying the floating phase, i.e., a region with $d=0$.  The
data generated with the infinite-volume DMRG at $0.3<B<0.4$ seem to
indicate $d<0$; this appears to be an artifact of the infinite-volume
DMRG, as we can see from the comparison of $E_{\rm DW}$ evaluated with
the finite- and infinite-volume DMRG at $B=0.3$, shown in
Fig.~\ref{fig:k=0.75,B=0.3,EDW1}.
\begin{figure}[tbp]
\begin{center}
\leavevmode
\includegraphics*[scale=\graphicscale]{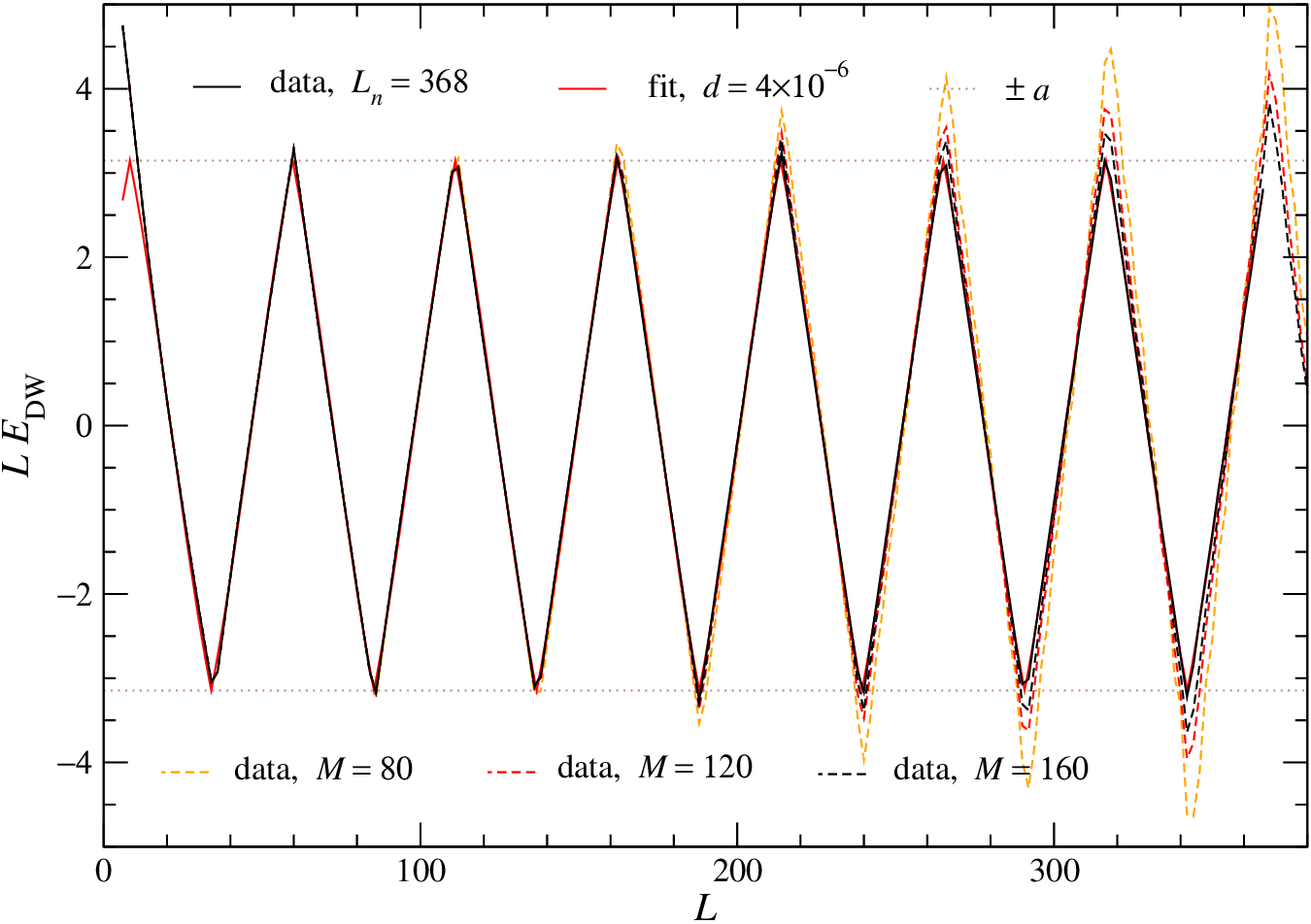}
\caption{Determinations of the domain-wall energy 
$E_{\rm DW}$, multiplied by $L$, vs.\ $L$, for $\kappa=0.75$ and
$B=0.3$: finite-volume DMRG at $M=80$, 120, 160; infinite-volume DMRG
at $L_n=368$, $M=80$; fit to Eq.\ (\ref{eq:DW-para}) of the $L_n=368$
data.}
\label{fig:k=0.75,B=0.3,EDW1}
\end{center}
\end{figure}

We must therefore resort to the resource-consuming finite-volume DMRG.
We checked in several instances that $M=80$ is sufficient to obtain
accurate results and that obtaining the data for all $L$s from a run
at a single $L_n$ is acceptable: in all cases, the determinations of
$d$ are well within the error quoted.

We show the results in Fig.~\ref{fig:k=0.75,d,EDW}.  The value of $d$
obtained from the finite-volume DMRG is consistent with zero up to
$B=0.425$, where we estimate $\xi\equiv1/d>10^4$.  The smoothness of
$d$ vs.\ $B$ suggest a higher-order, possibly Kosterliz-Thouless,
phase transition.  Note that $d$ is quite compatible with $1/\xi$ of
Fig.~\ref{fig:k=0.75,xi-SA} (apart from a normalization stemming from
the different definition of correlation length), and so is the
resulting $B_2=0.435(10)$.
\begin{figure}[tbp]
\begin{center}
\leavevmode
\includegraphics*[scale=\graphicscale]{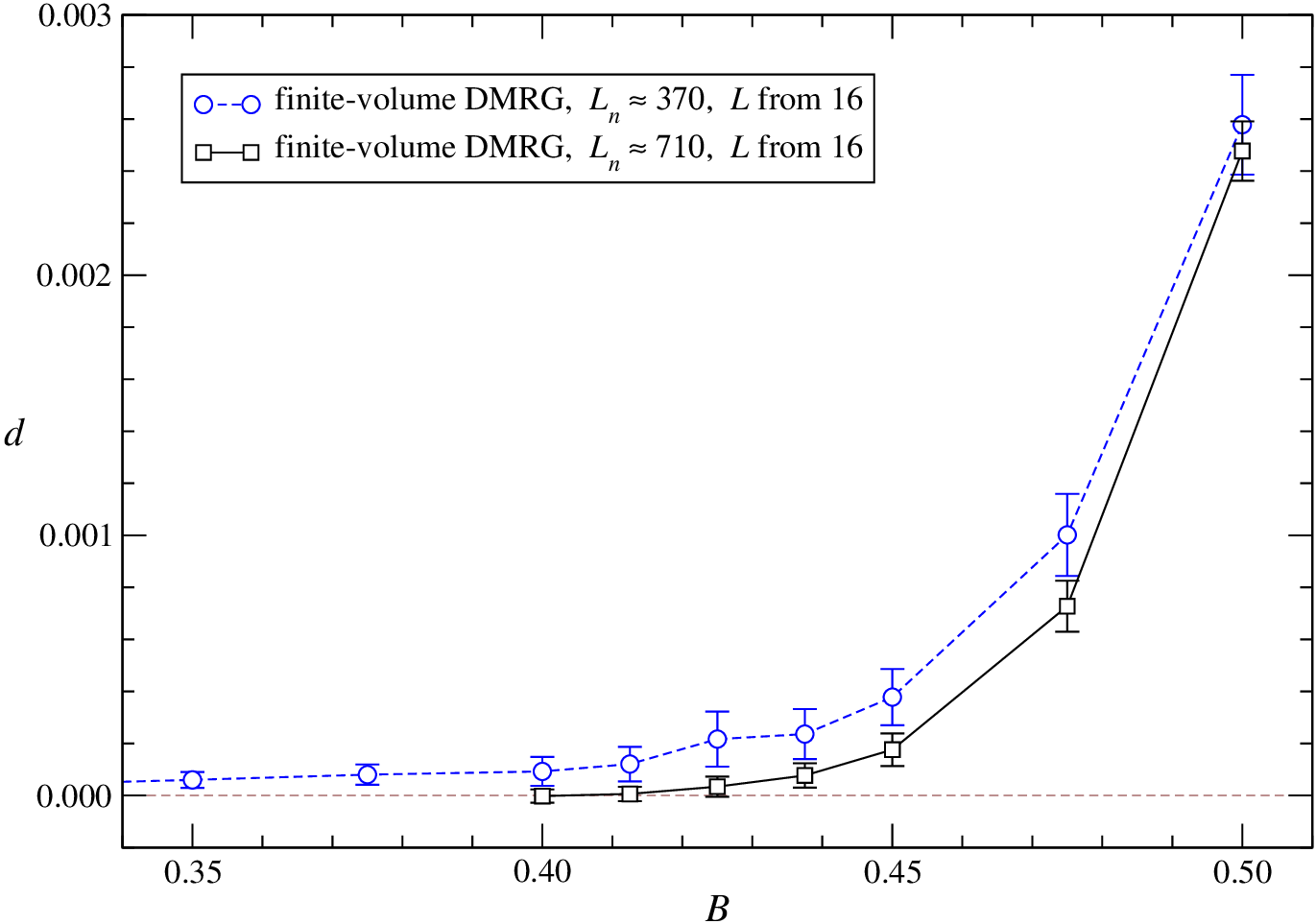}
\caption{The decay parameter $d$ of the domain-wall
energy $E_{\rm DW}$, computed with the finite-volume DMRG, vs.\ $B$,
for $\kappa=0.75$.}
\label{fig:k=0.75,d,EDW}
\end{center}
\end{figure}

The precise determination of the transition point of a
Kosterliz-Thouless phase transition is a notoriously difficult
problem: it is always possible that a system with a huge correlation
length is mistaken for a critical system\cite{Shirahata-Nakamura}.
Indeed, the correlation length is expected to diverge very rapidly
when the Kosterliz-Thouless critical coupling is approached; following
Ref.~\onlinecite{Hu}, we could conjecture a behavior $\xi \propto
\exp(b/(B-B_2))$ for $B\searrow B_2$.  A fit to the above form of the
data of Fig.\ \ref{fig:k=0.75,d,EDW} gives unstable results for
$B_2$, indicating that the determination of $B_2$ from $E_{\rm DW}$
should be treated with some caution.  On the other hand, a fit to the
data of Fig.~\ref{fig:k=0.75,xi-SA} gives results which are stable and
fully consistent with the estimate of $B_2$ from $S_A$ given above.

The two determinations of $B_2$ from $\Delta_L$ and $S_A$, obtained by
quite different methods, are accurate and in agreement with each other
(and with the less reliable determination from $E_{\rm DW}$); moreover,
for each method, we see no trend of $B_2$ decreasing with increasing
$L$ or $M$ (see especially Figs.~\ref{fig:k=0.75,B_i} and
\ref{fig:k=0.75,xi-SA}).  
We can conclude that, while it is possible that the errors on
$B_2$ are underestimated, it is difficult to believe that the
floating phase might disappear completely on larger systems.

\section{Phase diagram}
\label{sec:phase-diagram}

The study of the phase transitions at other values of $\kappa$s is
very similar to the one at $\kappa=0.75$ presented in
Sect.~\ref{sec:phases-k=0.75} and we can avoid repeating the details.
We selected for our analysis the values $\kappa=0.5$, 0.52, 0.55, 0.6,
0.75, 1.0, 1.25, 1.5, 2.0, 5.0.

At $\kappa=0.5$, the DMRG algorithm becomes inefficient at low $B$,
and we are unable to run at $B<0.01$.  We see no sign of a floating
phase: the curves $L\Delta_L(B,\kappa)$ vs.\ $B$ almost coincide for
$B<0.06$, and the intersections are very unstable.  Determining
$\xi$ by fitting $S_A$ with $c\equiv1$ fixed, we see no sign of
$\xi=\infty$, see Fig.~\ref{fig:k=0.5,xi-SA}; given the poor
convergence in $M$ for small $B$, we estimate $B_2<0.04$.
\begin{figure}[tbp]
\begin{center}
\leavevmode
\includegraphics*[scale=\graphicscale]{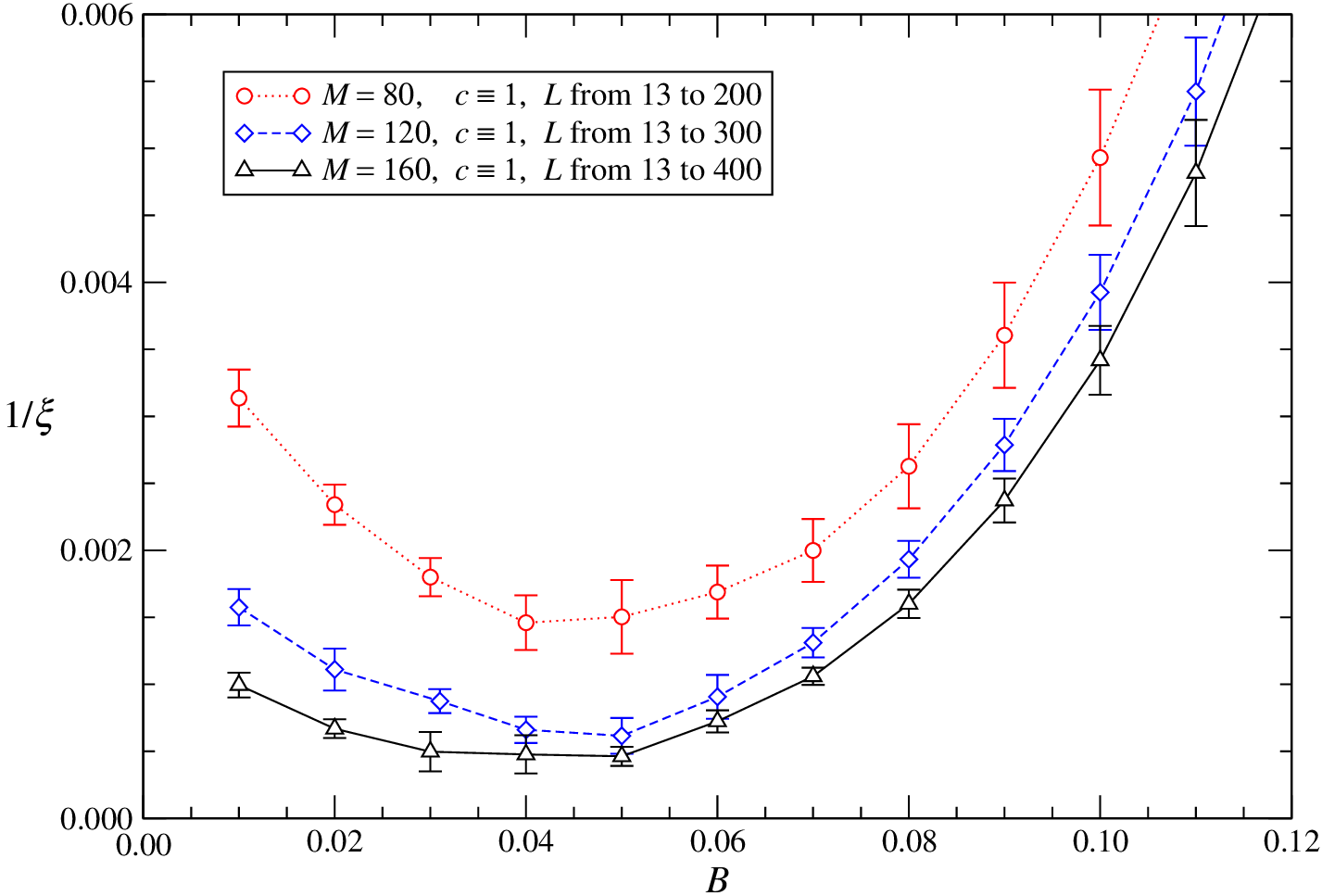}
\caption{The reciprocal correlation length $1/\xi$ (determined from
$S_A$) vs.\ $B$, for $\kappa=0.5$.}
\label{fig:k=0.5,xi-SA}
\end{center}
\end{figure}
The analysis of $E_{\rm DW}$ does not give precise results for $B_2$.
The behavior of the modulation parameter $k$ (see
Fig.~\ref{fig:k=0.5,k2-EDW}), which goes to a nonzero value as
$B\to0$, hints at the very peculiar nature of the multicritical point
at $\kappa=0.5$, $B=0\;$ \cite{Selke,Sen-Das}.
\begin{figure}[tbp]
\begin{center}
\leavevmode
\includegraphics*[scale=\graphicscale]{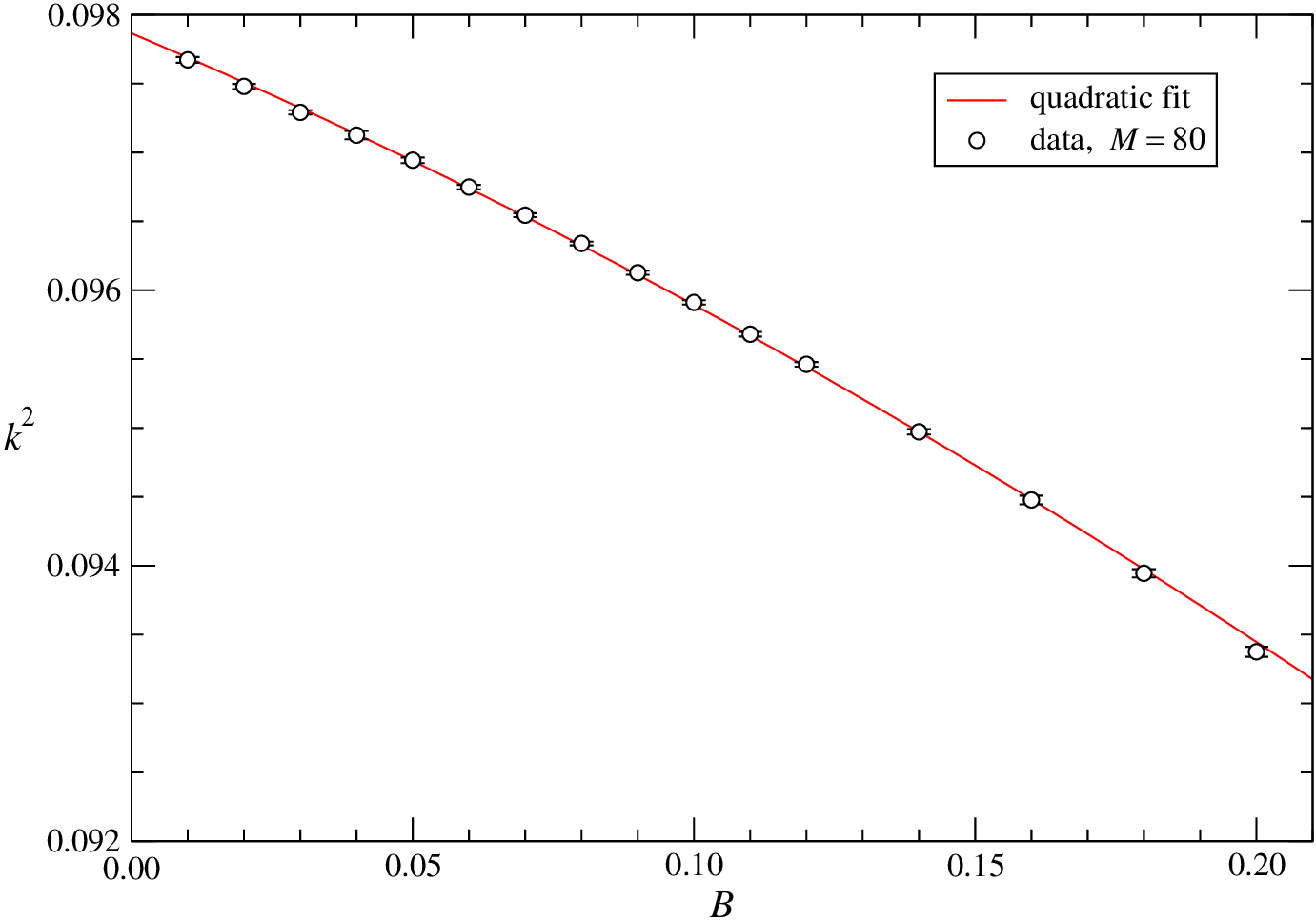}
\caption{The squared modulation parameter $k^2$ 
vs.\ $B$, for $\kappa=0.5$.}
\label{fig:k=0.5,k2-EDW}
\end{center}
\end{figure}

For $\kappa=0.52$ and 0.55, the quality of the determinations of $B_1$
and of the determination of $B_2$ from $S_A$ is similar to those at
$\kappa=0.75$; on the other hand, the analysis of $\Delta_L$ and
$E_{\rm DW}$ do not give precise results for $B_2$.  We present the
plot of $1/\xi$ for $\kappa=0.52$ in Fig.~\ref{fig:k=0.52,xi-SA}: 
the difference from Fig.~\ref{fig:k=0.5,xi-SA} is remarkable.
\begin{figure}[tbp]
\begin{center}
\leavevmode
\includegraphics*[scale=\graphicscale]{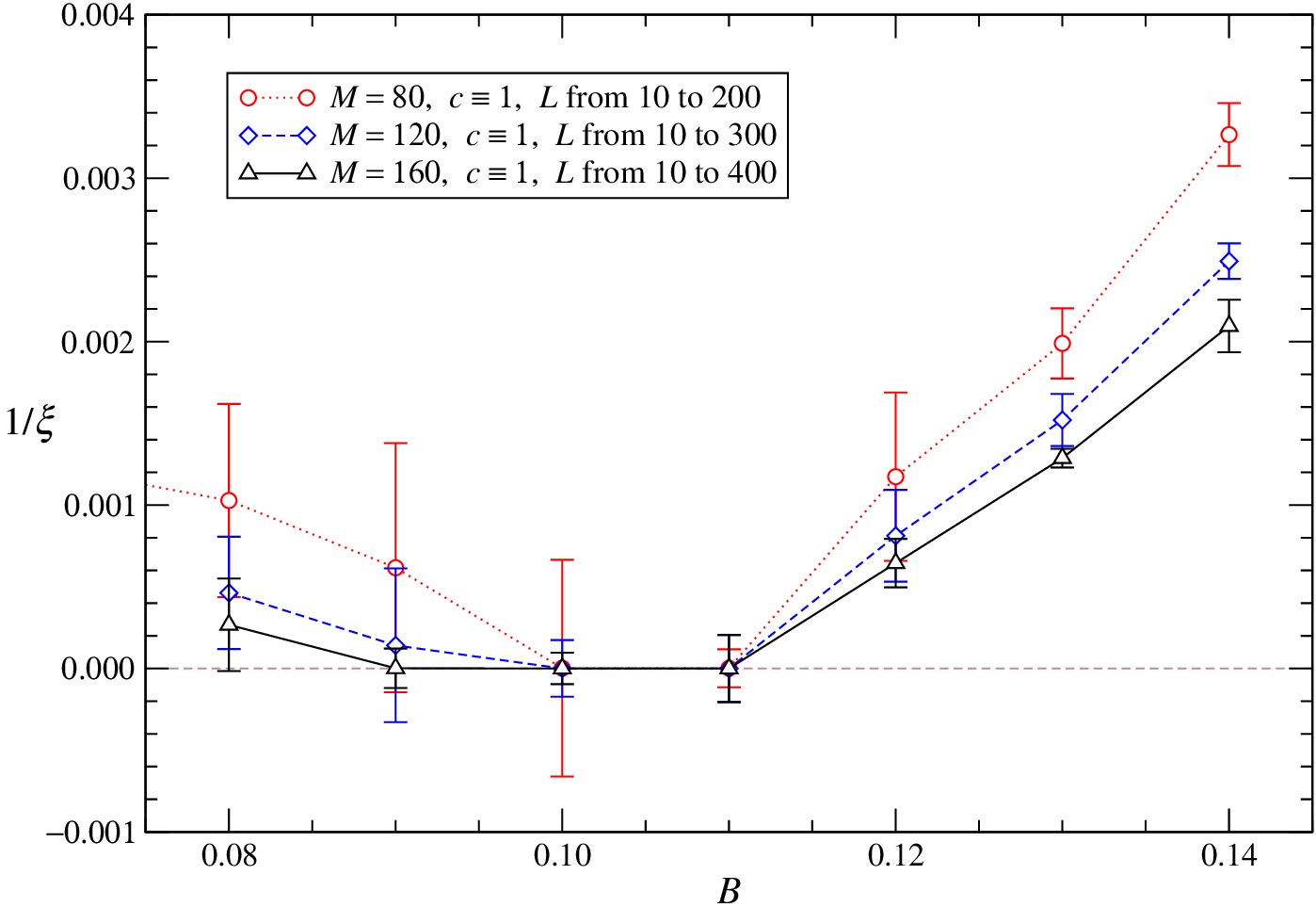}
\caption{The reciprocal correlation length $1/\xi$
(determined from $S_A$) vs.\ $B$, for $\kappa=0.52$.}
\label{fig:k=0.52,xi-SA}
\end{center}
\end{figure}

For $0.6\le\kappa\le1.5$, there are no relevant differences from the
case $\kappa=0.75$ described in Sect.~\ref{sec:phases-k=0.75}; we only
present the determinations of $B_1$ and $B_2$ in Table
\ref{tab:Btrans}.  For $\kappa=2$, the only difference is that 
$E_{\rm DW}$ is not fitted well by Eq.\ (\ref{eq:DW-para}) in the
floating and paramagnetic phases, and therefore the determination of
$B_2$ from $E_{\rm DW}$ is unreliable.

\begin{table*}
\begin{center}
\setlength\tabcolsep{6pt}
\begin{tabular}{|l|ll|lll|}
\hline
\multicolumn{1}{|c|}{$\kappa$} & \multicolumn{1}{c}{$B_1\ (E_{\rm DW})$} & 
\multicolumn{1}{c|}{$B_1\ (o)$} & \multicolumn{1}{c}{$B_2\ (\Delta_L)$} &
\multicolumn{1}{c}{$B_2\ (S_A)$} & \multicolumn{1}{c|}{$B_2\ (E_{\rm DW})$} \\
\hline
% k  | B1 (EDW)     B1 (o)    | B2 (gap)     B2 (SA)    B2 (EDW)
0.5  & 0          & 0         & $<0.06$    & $<0.04$  & $<0.08$     \\
0.52 & 0.0201(1)  & 0.021(1)  & 0.095(15)  & 0.115(5) & 0.12(3)     \\
0.55 & 0.0501(2)  & 0.052(2)  & 0.160(15)  & 0.175(5) & 0.18(2)     \\
0.6  & 0.1015(2)  & 0.103(2)  & 0.235(6)   & 0.25(1)  & 0.25(1)     \\
0.75 & 0.2574(2)  & 0.2575(5) & 0.424(3)   & 0.44(1)  & 0.425(10)   \\
1.0  & 0.5213(2)  & 0.522(2)  & 0.700(5)   & 0.72(1)  & 0.71(1)     \\
1.25 & 0.7867(2)  & 0.785(2)  & 0.972(4)   & 1.00(1)  & 0.98(1)     \\
1.5  & 1.0514(2)  & 1.045(5)  & 1.235(3)   & 1.26(1)  & 1.26(1)     \\
2.0  & 1.5775(2)  & 1.576(2)  & 1.756(6)   & 1.79(1)  & 1.79(3)     \\
5.0  & \multicolumn{1}{c}{---} & 4.667(3)  & \multicolumn{1}{c}{---} 
     & 4.88(1)  & \multicolumn{1}{c|}{---} \\
\hline 
\end{tabular}
\end{center}
\caption{Determinations of the transition fields $B_1$ and $B_2$ 
by different techniques.}
\label{tab:Btrans}
\end{table*}

In the case $\kappa=5$, the determination of $B_1$ and $B_2$ is rather
imprecise: $E_{\rm DW}$ is not fitted well by Eq.\ (\ref{eq:DW-anti})
in the antiphase, and it is fitted poorly by Eq.\ (\ref{eq:DW-para})
in the floating and paramagnetic phases; it is very hard to get
precise results from $\Delta_L$, since the modulation parameter is
very small ($k\lesssim0.01$ in the floating phase).  It is still
possible to estimate $B_1$ from $o$ and $B_2$ from $S_A$.

For all the values of $\kappa$ considered, the different
determinations of $B_1$ and $B_2$ are in substantial agreement with
each other: this is a strong argument supporting the reliability of
our results.  It should be noticed, however, that the determination of
$B_2$ from $\Delta_L$ is systematically lower than the determination
from $S_A$, possibly indicating that the error on the determination
from $\Delta_L$ reported in Table \ref{tab:Btrans} is underestimated.

We can beautifully summarize all the above results by noticing that
all the determinations of $B_1$ and $B_2$ are consistent with
\begin{equation}
B_1(\kappa) \cong 1.05 (\kappa-{\textstyle{1\over2}}), \quad
B_2(\kappa) \cong 1.05 \sqrt{(\kappa-{\textstyle{1\over2}})(\kappa-0.1)}.
\label{eq:B12-fit}
\end{equation}

\begin{figure}[tbp]
\begin{center}
\leavevmode
\includegraphics*[scale=\graphicscale]{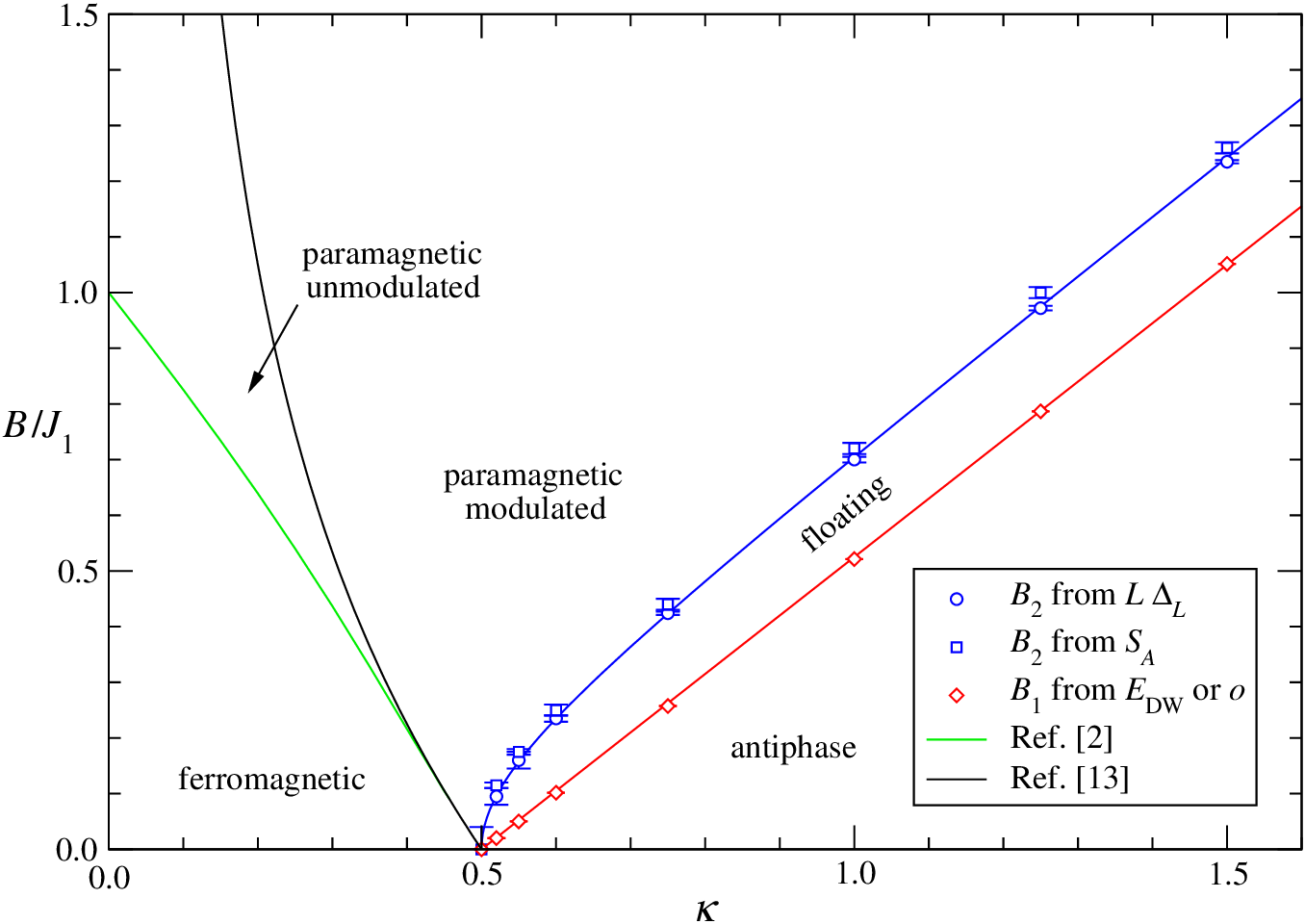}
\caption{Phase diagram in the $\kappa$--$B$ plane; the solid lines for
$\kappa\ge0.5$ correspond to Eq.\ (\ref{eq:B12-fit}).}
\label{fig:phase-diagram}
\end{center}
\end{figure}

Finally, we draw the phase diagram in the $\kappa$--$B$ plane in
Fig.~\ref{fig:phase-diagram}.  The region $\kappa<0.5$ was studied in
Ref.~\onlinecite{Beccaria:2005nz}; the critical line separating the
paramagnetic modulated and paramagnetic unmodulated phases is known
analytically\cite{Peschel-Emery}.  The data in the region $\kappa>0.5$
are taken from the present work; note that earlier
results\cite{Duxbury-Barber,Arizmendi-etal,Sen-etal,Guimaraes-etal}
provided only a qualitative picture of the phase diagram in this
region.

A very interesting question is whether the floating phase extends up
to $\kappa=\infty$ or it terminates at finite $\kappa$; we found that
the floating phase extends at least up to $\kappa=5$.

\section{Summary and conclusions}
\label{sec:conclusions}

We applied the DMRG algorithm to the study of the quantum transverse
ANNNI model in the region of high frustration ($\kappa>0.5$). 

We obtained clear evidence for the existence of a floating phase for
$\kappa>0.5$, extending at least up to $\kappa=5$.  The floating phase
is separated from the paramagnetic modulated phase by a high-order
(possibly Kosterliz-Thouless) critical line, ending at the
multicritical point ($\kappa=0.5$, $B=0$); the corresponding central
charge is $c=1$.  In Ref.\ \onlinecite{Chandra-Dasgupta}, the floating
phase was shown to have a finite extent at $\kappa=0.5$; our study
cannot exclude a floating phase of very small extent, i.e.,
$0<B_2(\kappa{=}0.5)\lesssim0.04$.

We obtained precise estimates for the critical points, verifying that
different methods give consistent results.  Simple and
accurate formulae for the two critical lines are reported in Eq.\ 
(\ref{eq:B12-fit}).

Very helpful discussions with Walter Selke and Pasquale Calabrese are
gratefully acknowledged.


\begin{thebibliography}{99}

\bibitem{Selke}
  W.~Selke,
  Phys.\ Rep.\ {\bf 170}, 213 (1998).

\bibitem{Beccaria:2005nz}
  M.~Beccaria, M.~Campostrini, and A.~Feo,
  Phys.\ Rev.\ {\bf B 73}, 052402 (2006).

\bibitem{Duxbury-Barber}
  Ph.~Duxbury and M.~Barber, 
  J.\ Phys.\ A: Math.\ Gen.\ {\bf 14}, L251 (1981).

\bibitem{Arizmendi-etal}
  C.~M.~Arizmendi, A.~H.~Rizzo, L.~N.~Epele, and C.~A.~Garcia Canal,
  Z.\ Phys.\ {\bf B~83}, 273 (1991).

\bibitem{Sen-etal}
   P.~Sen, S.~Chakraborty, S.~Dasgupta, and B.~K.~Chakrabarti,
  Z.\ Phys.\ {\bf B~88}, 333 (1992).

\bibitem{Guimaraes-etal}
  P.~R.~Colares Guimar\~aes, J.~A.~Plascak, F.~C.~S\`a Barreto, and
  J.~Florencio,
  Phys.\ Rev.\ {\bf B66}, 064413 (2002).

\bibitem{Chandra-Dasgupta}
  A.~K.~Chandra and S.~Dasgupta,
  Phys.\ Rev.\ {\bf E~75}, 021105 (2007).

\bibitem{Calabrese:2004eu}
  P.~Calabrese and J.~L.~Cardy,
  J.\ Stat.\ Mech.\ {\bf 0406}, P002 (2004).

\bibitem{Derian-Gendiar-Nishino}
  R.~Derian, A.~Gendiar, and T.~Nishino,
  J.~Phys.\ Soc.\ Jpn.\ {\bf 75}, 114001 (2006).

\bibitem{Schollwoeck-review}
  U.~Schollw\"ock,
  Rev.\ Mod.\ Phys.\ {\bf 77}, 259 (2005); see also\\
  {\tt http://quattro.phys.sci.kobe-u.ac.jp/dmrg.html}

\bibitem{ARPACK}
  {\tt http://www.caam.rice.edu/software/ARPACK/}

\bibitem{Shirahata-Nakamura}
  T.~Shirahata and T.~Nakamura,
  Phys.\ Rev.\ {\bf B 65}, 024402 (2001).

\bibitem{Hu}
  Xiao Hu, Prog.\ Theor.\ Phys.\ {\bf 89}, 545 (1993); 
  Xiao Hu, J.\ Phys.\ A: Math.\ Gen.\ {\bf 27}, 2313 (1994).

\bibitem{Sen-Das}
  P.~Sen and P.~K.~Das,
  ``Dynamical frustration in ANNNI model and annealing'',
  in ``Quantum Annealing and Related Optimization Methods'', 
  A.~Das and B.~K.~Chakrabarti, eds., 
  Lecture Notes in Physics 679, Springer, New York, 2005.

\bibitem{Peschel-Emery}
  I.~Peschel and V.~J.~Emery, Z.\ Phys.\ {\bf B43}, 241 (1981).

\end{thebibliography}
\end{document}